\documentclass[useAMS,usenatbib]{mn2e}
\usepackage{graphicx,epsfig}

\def\msun{\hbox{M$_{\odot}$}}

\def\lsun{\hbox{L$_{\odot}$}}
\def\msunyr{\mbox{\,${\rm M_{\odot}\, yr^{-1}}$}}
\def\mdot{\dot M}

\def\degs{\ifmmode ^{\circ}\else$^{\circ}$\fi}
\def\amin{\ifmmode ^{\prime}\else$^{\prime}$\fi}
\def\asec{\ifmmode ^{\prime\prime}\else$^{\prime\prime}$\fi}
\def\fss{\hbox{$.\!\!^{\rm s}$}}        

\def\farcs{\hbox{$.\!\!^{\prime\prime}$}}  
\def\h{\hbox{$^{\rm h}$}}
\def\m{\hbox{$^{\rm m}$}}

\def\EE#1{\times 10^{#1}}

\def\cm{\mbox{\,cm}}

\def\cm3{\mbox{\,cm$^{-3}$}}

\def\ergs{\mbox{\,ergs~s$^{-1}$}}
\def\ergshz{\mbox{~ergs~s$^{-1}$~Hz$^{-1}$}}
\def\whz{\mbox{~W~$~Hz^{-1}$}}
\def\kms{\mbox{\,km s$^{-1}$}}

\def\mJybeam{~mJy~beam$^{-1}$}
\def\muJybeam{~$\mu$Jy~beam$^{-1}$}
\def\lsim{\!\!\!\phantom{\le}\smash{\buildrel{}\over
 {\lower2.5dd\hbox{$\buildrel{\lower2dd\hbox{$\displaystyle<$}}\over
                                 \sim$}}}\,\,}
\def\gsim{\!\!\!\phantom{\ge}\smash{\buildrel{}\over
{\lower2.5dd\hbox{$\buildrel{\lower2dd\hbox{$\displaystyle>$}}\over
                               \sim$}}}\,\,}

\title[The continuum radio emission from the Sy~1.5 galaxy NGC~5033]{Radio emission from the Sy~1.5 galaxy NGC~5033}

\author[M. A. P\'erez-Torres and A. Alberdi]{M. A. P\'erez-Torres$^{1}$\thanks{E-mail:torres@iaa.es (MAPT); antxon@iaa.es (AA)} and A. Alberdi$^{1}$ \\
$^{1}$Instituto de Astrof\'{\i}sica de Andaluc\'{\i}a, CSIC, 
Apdo. Correos 3004, E-18080 Granada, Spain\\
}

\begin{document}

\date{Accepted 2007 May 4. Received  2007 May 2; in original form 2007 March 23}

\pagerange{\pageref{firstpage}--\pageref{lastpage}} \pubyear{YYYY}

\maketitle

\label{firstpage}

\begin{abstract}
We present new continuum VLA observations of the nearby Sy~1.5 galaxy
NGC~5033, made at 4.9 and 8.4 GHz on 8 April 2003.  Combined with VLA
archival observations at 1.4 and 4.9~GHz made on 7 August 1993, 29
August 1999, and 31 October 1999, we sample the galaxy radio emission
at scales ranging from the nuclear regions ($\lsim$100~pc) to the
outer regions of the disk ($\sim 40$~kpc).  The high-resolution VLA
images show a core-jet structure for the Sy~1.5 nucleus. While the
core has a moderately steep non-thermal radio spectrum ($S_{\nu}
\propto \nu^\alpha; \alpha_{1.5}^{4.9} \approx -0.4$), the inner kpc
region shows a steeper spectrum ($\alpha_{1.5}^{8.4} \approx -0.9$).
This latter spectrum is typical of galaxies where energy losses are
high, indicating that the escape rate of cosmic ray electrons in
NGC~5033 is low.  The nucleus contributes little to the total 1.4~GHz
radio power of NGC~5033 and, based on the radio to far-infrared (FIR)
relation, it appears that the radio and far-infrared emission from
NGC~5033 are dominated by a starburst that during the last 10~Myr
produced stars at a rate of 2.8\msunyr\, yielding a supernova (type
Ib/c and II) rate of 0.045\,yr$^{-1}$. This supernova rate corresponds
to about 1 SN event every 22 yr.  Finally, from our deep 8.4~GHz VLA-D
image, we suggest the existence of a radio spur in NGC~5033, which
could have been due to a hot superbubble formed as a consequence of
sequential supernova explosions occurring during the lifetime of a
giant molecular cloud.
\end{abstract}

\begin{keywords}
 Galaxies: individual: NGC\,5033 -
 Galaxies: Seyfert, starburst - 
 Radio continuum: general -- 
 Radiation mechanisms: non-thermal 
\end{keywords}

\section{Introduction}\label{sec,intro}

Radio emission from normal galaxies is  dominated by
synchrotron emission from relativistic cosmic ray electrons (CRE) at
low frequencies ($\lsim$30~GHz), while
at high frequencies ($\nu \gsim$30~GHz) the dominant emitting
mechanism is thermal free-free emission from ionized gas at
temperatures of $\sim\,10^4~$K.

With the advent of the Infrared Astronomical Satellite (IRAS,
Neugebauer et al. 1984), a sample of $\approx 20,000$ galaxies
complete to 0.5 Jy in the 60 $\mu$m band was detected, the majority of
which had not been previously cataloged.  While the observed FIR
luminosity from most of those galaxies can be explained by ongoing
star formation and starburst activity, some of the most luminous
infrared galaxies may host an active galactic nucleus (AGN).  Nearly
all of the radio emission at wavelengths longer than a few cm from
such galaxies is synchrotron radiation from relativistic electrons and
free-free emission from H~II regions.  In turn, this synchrotron radio
emission is a direct probe of the recent massive starforming activity,
since only stars more massive than about 8\msun result in the Type
Ib/c and II supernovae, which are thought to accelerate most of the
relativistic electrons that are responsible for the observed
synchrotron radio emission.  This results in an extremely good
correlation between radio and far-infrared emission among galaxies
(e.g., \citealt{harwit75}, \citealt{condon91a},
\citealt{condon91b}). If an AGN is present, then the associated radio
continuum, in excess of the level expected from the star formation,
may be detected and, by means of high-resolution observations, its
contribution to the total radio emission subtracted.

Synchrotron radio emission also carries relevant information on the
magnetic field and the particle feeding, or re-acceleration mechanisms
present in galaxies.  Since the synchrotron energy loss rate of
relativistic electrons, $dE/dt$, varies with the particle energy
squared, $E^2$, the typical lifetime of a relativistic electron,
subject to radiative synchrotron losses is $t_{\rm syn} = E/(dE/dt)
\propto E^{-1}$.  Further, the critical frequency at which an electron
radiates most of its synchrotron emission is $\nu_{\rm c} \propto
E^2$, and hence $dE/dt \propto \nu_{\rm c}$, and $t_{\rm syn} \propto
\nu^{-1/2}_{\rm c}$.  Therefore, regions with little ongoing
relativistic particle re-supply, or re-acceleration have steeper
synchrotron spectra, which may be revealed by multi-frequency radio
interferometry observations.  The sites and length scales where the
spectral evolution is happening can also be revealed by
high-resolution radio interferometry observations, which tell us where
CRE are accelerated.  

High-resolution radio observations also allow the detailed study of
both thermal and non-thermal emission in nearby, normal galaxies
(e.g. \citealt{condon92}). Those observations are needed, as the
spatially integrated nonthermal spectral index is in general a poor
diagnostic for the type of propagation or the importance of energy
losses.  On the contrary, spatially resolved radio data for the haloes
of galaxies allow us to draw firm conclusions. For example, the steepening
of the spectrum away from the disk is an indication that
synchrotron and inverse Compton losses are taking place during the
propagation of cosmic ray electrons in the halo \citep{lisenfeld00}.

NGC~5033 is a nearby ($D$=13~Mpc) spiral galaxy that exhibits strong
CO emission out to at least $R\sim50$\farcs, and contains a Seyfert
1.5 nucleus \citep{ho97}.  NGC~5033 also displays strong far infrared
luminosity ($L_{\rm [8-1000 \mu~m]} = 1.20\EE{10}$\lsun\,;
\citealt{sanders03}), and has been host to at least three supernovae
in the last 60 yr (SN~1950C, SN~1985L, and SN~2001gd).  Here, we
present and discuss continuum VLA-D observations of NGC~5033 at 1.4,
4.9 and 8.4 GHz taken on April 8, 2003, complemented with archival VLA
data taken on August 29, 1999 (1.4 GHz, A-configuration) and October
31, 1999 (4.9 GHz, B-configuration), respectively.  The resolutions
yielded by the different arrays and frequencies allow us to trace the
synchrotron emission not only in the innermost regions of the galaxy,
but also in the inner and outer disk, and the galactic halo.

The paper is organized as follows: we report on the radio observations
in Sect.~\ref{sec,obs}; we present our results in
Sect.~\ref{sec,results} and discuss them in
Sect.~\ref{sec,discussion}.  We summarize our main conclusions in
Sect.~\ref{summary}.  We assume throughout the paper a distance of
13~Mpc to the host galaxy of SN~2001gd, NGC~5033, based on its
redshift ($z=0.002839$; \citealt{falco99}) and assumed values of $H_0
= 65\kms$~Mpc$^{-1}$ and $q_0$=0.  At the distance of NGC~5033
1\arcsec\ corresponds to a linear size of 64~pc.

\begin{figure*}
\begin{center}
 \includegraphics[angle=0,width=\textwidth]{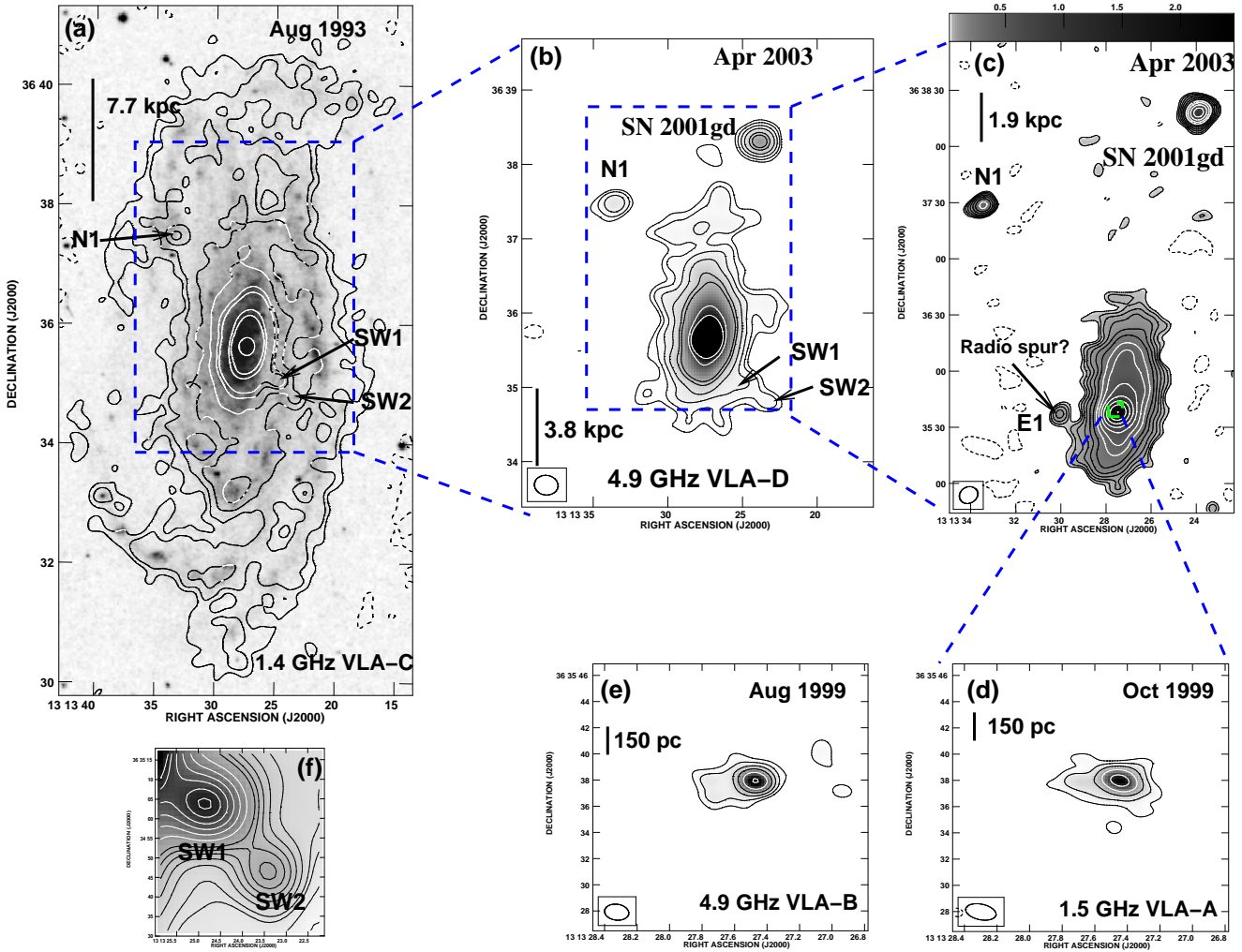}
\caption{ {\bf(a)} Contours of 1.4~GHz observations made on 7 August 1993
  with the VLA-C, overlaid on a Digitized Sky Survey (DDS2) Blue plate
  of NGC~5033; {\bf (b,c)} total intensity radio images of NGC\,5033
  obtained from 4.9 and 8.4~GHz VLA-D observations on 8 April 2003;
  {\bf (d,e)} 1.4~GHz VLA-A observations on 29 August 1999 and 4.9~GHz VLA-B
  observations on 31 October 1999; {\bf (f)} blow-up of the 1.4~GHz VLA-C
  image, around the SW1 and SW2 regions.  VLA contours are drawn at
  (-3,3,$3\,\sqrt{2}$,9,...)$\times$ the off-source rms of each map,
  which are of 93\,$\mu$Jy beam$^{-1}$ (1.4~GHz VLA-C), 45\,$\mu$Jy
  beam$^{-1}$ (4.9~GHz VLA-D), 19 \,$\mu$Jy beam$^{-1}$ (8.4~GHz
  VLA-D), and of 170 and 70 \,$\mu$Jy beam$^{-1}$ at 1.4 and 4.9 GHz
  (panels d and e), respectively.  The peaks of brightness are of 31.8,
  10.6, and 2.4~mJy beam$^{-1}$ at 1.4, 4.9, and 8.4~GHz
  (top panels), and of 5.7 and 3.4~mJy beam$^{-1}$ at 1.5 and 4.9~GHz
  (panels d and e).  The principal dimensions (major axis
  $\times$ minor axis, position angle) of the restoring beams are:
  17\farcs3 $\times$15\farcs4, 83\degr (a); 19\farcs1
  $\times$15\farcs9, 83\degr (b); 10\farcs1 $\times$8\farcs7,
  -61\degr (c); 2\farcs3 $\times$1\farcs2, 78\degr (d); 
  1\farcs8 $\times$1\farcs2, 82\degr (e).  }
\label{fig,vla}
\end{center}
\end{figure*}

\begin{table*}
 \caption[]{VLA sources within the galaxy NGC~5033}
  \label{tab,vla}
$$
 \begin{array}{lllrrrrrr}
   \hline\noalign{\smallskip}
{\rm Source} & 
\multicolumn{1}{c}{\alpha(J2000.0)} & \multicolumn{1}{c}{\delta(J2000.0)} &  
\multicolumn{1}{c}{S^P_{1.4}} & \multicolumn{1}{c}{S^I_{1.4}} & 
\multicolumn{1}{c}{S^P_{4.9}} & \multicolumn{1}{c}{S^I_{4.9}} & 
\multicolumn{1}{c}{S^P_{8.4}} & \multicolumn{1}{c}{S^I_{8.4}} \\
              &                  
              &                 &  
\multicolumn{1}{c}{\rm (mJy/b)} & \multicolumn{1}{c}{\rm (mJy)} &
\multicolumn{1}{c}{\rm (mJy/b)} & \multicolumn{1}{c}{\rm (mJy)} &
\multicolumn{1}{c}{\rm (mJy/b)} & \multicolumn{1}{c}{\rm (mJy)} \\
   \noalign{\smallskip}
   \hline\noalign{\smallskip}
{\rm NGC~5033*^\dagger} & \rm 13\h13\m27\fss4430 & \rm 36\degs35\amin37\farcs880 
            &  5.7\pm0.15  & 12.73\pm0.30 
            &  3.83\pm0.02 &  6.25\pm0.33
            & \cdots       & \cdots            \\
{\rm NGC~5033^\ddagger}  & \cdots       & \cdots
            & 31.9\pm0.09  & 355.41\pm1.64
            & 10.6\pm0.03  &  45.46\pm0.23
            &  2.4\pm0.02  &  11.72\pm0.12 \\
{\rm N1^a}     & 13\h13\m33\fss4517 &  36\degs37\amin29\farcs116 
   & 1.63\pm0.09 & 3.01\pm0.24 & 0.66\pm0.03 & 0.70\pm0.05 & 0.36\pm0.02 &0.29\pm0.02  \\
{\rm SW1^a}   &  13\h13\m24\fss862    &  36\degs35\amin04\farcs48  
          & 4.46\pm0.09      & 9.44\pm0.28 & 0.47\pm0.03  & 2.17\pm0.15 &\cdots &\cdots    \\
{\rm SW2^b}      & 13\h13\m23\fss995     &  36\degs34\amin38\farcs95  
          & 1.85\pm0.09      & 4.33\pm0.29 & 0.32\pm0.03  & 0.53\pm0.07 &\cdots &\cdots     \\
{\rm E1^a}       &  13\h13\m30\fss435    &  36\degs35\amin37\farcs130 
          & \cdots    &\cdots &\cdots &\cdots &0.12\pm0.02 & 0.12\pm0.02 \\
\hline\noalign{\smallskip}
 \end{array}
$$
\begin{list}{}{}
\item[] { $^\dagger$ The coordinates correspond to the putative
  nucleus of NGC~5033, dubbed NGC~5033$^*$, as seen with the VLA-A and
  VLA-B at 1.4~GHz (October 1999) and 4.9 GHz (August 199),
  respectively.  $S_{\nu}^P$ and $S_{\nu}^I$ are the peak and
  integrated flux densities for each given component, at a frequency
  $\nu$.  $^\ddagger$ Peak and integrated flux densities for NGC~5033,
  as seen with the VLA-C and VLA-D at 1.4 (August 1993), 4.9 (April
  2003), and 8.4 GHz (April 2003).  $^a$ Source tentatively identified
  as background source by \citet{ho01}; $^b$ Source position
  coincident with an H~II region \citep{evans96}. }
\end{list}
\end{table*}

\section{Observations}
\label{sec,obs}

We reanalysed archival VLA data on NGC~5033 obtained under program
AU0079 on 1999 August 29 (1.4 GHz, A-configuration) and 1999 October
31 (4.9 GHz, B-configuration), respectively. These data sets provide
very similar angular resolution, since the increase in angular
resolution from 1.4 to 4.9~GHz is essentially compensated by a
decrease in the longest antenna separation, which is approximately a
factor of three between the A and B configurations of the VLA. In both
observing epochs, the VLA recorded both senses of circular
polarization. Each frequency band was split into two intermediate
frequencies (IFs), of 50 MHz bandwidth each.  3C~286 was used to set
the absolute VLA flux density scale, and J1310+323 was used as the
phase calibrator at both frequencies.

We also show unpublished archival 1.4 GHz continuum VLA data obtained under
program AP270 on 1993 August 7, when the VLA was in C-configuration.
The VLA recorded only the left sense of circular polarization, and the
observations were made in spectral line mode, since the aim was to
image the 21-cm hydrogen line emission of NGC~5033. The 6.2-MHz IF at
1.4~GHz was split into 63 spectral channels, each of 97.66~kHz
bandwidth.  For the purposes of imaging the continuum radio emission
at 1.4~GHz, we used the channel-0 data, which contains the inner 75\%
of the line data.  3C~286 was used to set the absolute VLA flux
density scale, and the source 1323+321 was used as the phase
calibrator.  While the total synthesized bandwidth was small, the
large time on-source (about 5.5~hr) and excellent uv-coverage 
allowed us to obtain the deepest 21~cm image ever of the Sy~1.5 galaxy
NGC~5033 (see top left panel in Fig.~\ref{fig,vla}).

Finally, we show observations at 4.9 and 8.4~GHz of SN~2001gd
and its host galaxy NGC~5033, made on 2003 April 8 with the VLA in
D-configuration, as part of our VLBI observing program to monitor the
angular expansion of SN~2001gd \citep{mapt05}.
The VLA recorded both senses of circular polarization, and each
frequency band was split into two intermediate frequencies (IFs), of
50 MHz bandwidth each.  We used 3C~286 to set the absolute VLA flux
density scale at both frequencies.  We used J1310+323 as the phase
calibrator at 4.9~GHz, while J1317+34 was used as the phase calibrator
at 8.4~GHz.

For all data sets, we used standard calibration and hybrid mapping
techniques within the Astronomical Image Processing System ({\it
  AIPS}) to obtain the images shown in Figure~\ref{fig,vla}.

\section{Results}
\label{sec,results}

Figure~\ref{fig,vla} and Table~\ref{tab,vla} summarize our
results for the total intensity radio images of NGC~5033, obtained
with the VLA.  The flux density errors given for the VLA measurements
in Table~\ref{tab,vla} represent one statistical standard
deviation, and are a combination of the off-source rms in the image and
a fractional error, $\epsilon$, included to account for the inaccuracy
of VLA flux density calibration and possible deviations of the primary
calibrator from an absolute flux density scale. The final errors,
$\sigma_f$, as listed in Table~\ref{tab,vla}, are taken as
$\sigma_f^2 = (\epsilon\,S_0)^2 + \sigma_0^2$ where $S_0$ is the
measured flux density, $\sigma_0$ is the off-source rms at a given
frequency, and $\epsilon$=0.05 at 1.4~GHz, and 0.02 at 4.9~GHz and
8.4~GHz, respectively. We note that the total quoted flux density
errors are dominated by the uncertainties in the VLA calibration
($\epsilon$).  The positions of the peaks of brightness for all the
VLA images coincide among them within the errors, and correspond to
the nucleus of NGC~5033.

The 1.4~GHz VLA-C image (panel a in Fig. 1) shows the whole disk of
NGC~5033, with an angular size of about
10\amin$\times$\,5\amin\ ($\simeq 38\,{\rm kpc }\times\,19$~kpc).  The
low-surface brightness, extended emission clearly follows the optical
emission from the spiral arms of the galaxy, and most of the local
maxima of both optical and radio emission are spatially coincident
with H~II regions \citep{evans96}. The total flux density above the
lowest drawn contour is about 355~mJy, of which $\sim$58~mJy are
within a circumnuclear region of 30\asec$\times$30\asec ($\approx
1.8\times 1.8$~kpc$^2$).  We also detected three strong sources
embedded within the extended 1.4 GHz emission of the disk, which we
dubbed N1, SW1, and SW2 in panels (a) and (f), and whose nature is
discussed in section \ref{sec,discussion}.

The extension of the 4.9~GHz VLA-D total intensity emission of
NGC~5033 (panel b) is about 230\asec$\times$\,130\asec ($\simeq
14.7\,{\rm kpc }\times\,8.3$~kpc), significantly smaller than the
1.4~GHz radio emitting region. Since the 1.4~GHz VLA-C observations
have a very similar resolution to our 4.9~GHz VLA-D observations, this
finding indicates that the external regions of the disk are partially
resolved at this frequency.  The total flux density of the nuclear
regions of NGC~5033 above the lowest drawn contour is
$\simeq$45.3~mJy.  Note also that the extended galactic emission at
4.9~GHz around SN~2001gd is resolved out.  Note that there is no
contamination of background synchrotron emission around the position
of SN~2001gd, which is clearly detected at 4.9 GHz.
We note that N1, SW1, and SW2 are also detected at this frequency, despite
the 10 years elapsed between the 1.4  and 4.9 GHz VLA observations.

The extension of the 8.4~GHz VLA-D emission from the nuclear regions of
NGC~5033 (panel c) is about 110\asec$\times$\,50\asec ($\simeq
6.9\,{\rm kpc }\times\,3.2$~kpc), a size similar to that of the
4.9~GHz radio emitting region.  The total flux density for the nuclear
region of NGC~5033 above the lowest drawn contour is $\simeq$11.7~mJy.
Note that component N1, as well as the supernova SN~2001gd, are clearly
detected outside the main bulk of the nuclear radio emitting region.
Note also that a new component, E1, is detected close to the
inner regions of the disk, while components SW1 and SW2 are not
detected above a limiting flux density of 57$\mu$Jy/b

Panels (d) and (e) of Fig.~\ref{fig,vla} correspond to radio images of
the nuclear regions of NGC\,5033, obtained at 1.5 and 4.9~GHz from
archival observations made with the VLA in A- and B-configuration,
respectively.  The peaks of brightness are of 5.7 and 3.4~mJy
beam$^{-1}$ at 1.5 and 4.9~GHz, respectively, and correspond to the
emission from the core. A jet pointing eastwards from the core is also
seen at both frequencies. The different position angle of the jet, as
obtained at 1.4 GHz with respect to 5.0 GHz cannot be fully accounted
for by the different synthesized beam at each of the two frequencies.
In fact, while the 1.4~GHz VLA data displays a very well collimated
jet, the 4.9~GHz VLA data shows a hint for a double-layer jet
structure, which is very suggestive of the spine/shear layer model
proposed by \citet{laing02} to explain the synchrotron radio emission
of Fanaroff-Riley radio galaxies.  We also note that, while the
position angle of the upper layer of the 4.9~GHz jet coincides well
with that seen at 1.4~GHz, the bottom layer of the 4.9~GHz jet is not
seen at 1.4~GHz. This could point to opacity effects at the base of
the jet playing a relevant role, which could be responsible for
shaping the position angles seen at 1.4 and 4.9 GHz.

\begin{table}
 \caption[]{1.4~GHz Background sources$^\dagger$}
  \label{tab,back}
$$
 \begin{array}{ccrr}
   \hline\noalign{\smallskip}
  \alpha(J2000.0) & \delta(J2000.0) & S^P_{1.4} & S^I_{1.4} \\
                  &                 & {\rm (mJy/b)} & {\rm (mJy)}  \\ 
  \noalign{\smallskip}
   \hline\noalign{\smallskip}
13\h 11\m 38\fss347 &  36\degs 41\amin 22\farcs20 &   0.73 &   0.74 \\ 
13\h 11\m 46\fss609 &  36\degs 20\amin 12\farcs49 &   3.29 &   4.45 \\	
13\h 11\m 48\fss016 &  36\degs 19\amin 57\farcs67 &  ^{1}4.77 &   7.06 \\	
13\h 11\m 55\fss739 &  36\degs 36\amin 06\farcs44 &   0.57 &   0.59 \\	
13\h 12\m 04\fss680 &  36\degs 23\amin 18\farcs17 &  ^{2}15.27 &  15.87 \\	
13\h 12\m 12\fss688 &  36\degs 13\amin 13\farcs92 &   4.70 &   7.55 \\	
13\h 12\m 14\fss370 &  36\degs 49\amin 01\farcs94 &   1.15 &   1.25 \\	
13\h 12\m 15\fss339 &  36\degs 18\amin 01\farcs52 &   1.08 &   1.39 \\	
13\h 12\m 24\fss732 &  36\degs 40\amin 51\farcs28 &   6.54 &   7.32 \\	
13\h 12\m 25\fss069 &  36\degs 37\amin 27\farcs05 &  ^{2}2.17 &   2.29 \\	
13\h 12\m 25\fss575 &  36\degs 23\amin 45\farcs68 &   0.76 &   0.92 \\	
13\h 12\m 25\fss608 &  36\degs 43\amin 26\farcs39 &   0.57 &   0.61 \\	
13\h 12\m 26\fss132 &  36\degs 23\amin 21\farcs07 &   1.31 &   1.61 \\	
13\h 12\m 26\fss534 &  36\degs 26\amin 25\farcs30 &   0.47 &   0.37 \\	
13\h 12\m 33\fss178 &  36\degs 12\amin 55\farcs26 &   1.32 &   1.65 \\	
13\h 12\m 34\fss779 &  36\degs 57\amin 18\farcs14 &   0.82 &   0.91 \\	
13\h 12\m 37\fss528 &  36\degs 13\amin 57\farcs09 &   0.81 &   1.55 \\	
13\h 12\m 40\fss263 &  36\degs 48\amin 29\farcs27 &   0.49 &   0.77 \\	
13\h 12\m 51\fss637 &  36\degs 11\amin 05\farcs08 &   0.72 &   0.87 \\	
13\h 12\m 57\fss602 &  36\degs 47\amin 42\farcs15 &   0.58 &   0.61 \\	
13\h 13\m 00\fss943 &  36\degs 25\amin 06\farcs26 &   0.54 &   0.64 \\	
13\h 13\m 01\fss709 &  36\degs 48\amin 33\farcs70 &   0.94 &   1.31 \\	
13\h 13\m 01\fss878 &  36\degs 47\amin 40\farcs72 &   1.10 &   1.24 \\	
13\h 13\m 02\fss086 &  36\degs 39\amin 43\farcs77 &   1.32 &   1.88 \\	
13\h 13\m 03\fss766 &  36\degs 33\amin 29\farcs96 &   5.16 &   4.71 \\	
13\h 13\m 04\fss877 &  36\degs 30\amin 05\farcs62 &   0.70 &   0.55 \\	
13\h 13\m 10\fss193 &  36\degs 34\amin 42\farcs07 &   ^{3}4.88 &   4.17 \\	
13\h 13\m 15\fss991 &  36\degs 24\amin 46\farcs22 &   3.15 &   5.14 \\	
13\h 13\m 16\fss664 &  36\degs 25\amin 06\farcs13 &   1.06 &   0.76 \\	
13\h 13\m 19\fss228 &  36\degs 39\amin 40\farcs95 &   0.49 &   2.57 \\	
13\h 13\m 22\fss943 &  36\degs 40\amin 05\farcs27 &   0.55 &   4.87 \\	
13\h 13\m 23\fss156 &  36\degs 34\amin 41\farcs93 &   1.19 &   3.08 \\	
13\h 13\m 23\fss495 &  36\degs 55\amin 24\farcs94 &   2.43 &   2.47 \\	
13\h 13\m 27\fss121 &  36\degs 46\amin 03\farcs38 &   0.72 &   0.48 \\	
13\h 13\m 39\fss284 &  36\degs 50\amin 20\farcs10 &   ^{2}2.79 &   3.28 \\	
13\h 13\m 49\fss120 &  36\degs 49\amin 26\farcs25 &   0.50 &   1.09 \\	
13\h 13\m 57\fss673 &  36\degs 27\amin 40\farcs99 &   0.53 &   0.52 \\	
13\h 14\m 07\fss458 &  36\degs 42\amin 08\farcs60 &   0.61 &   1.02 \\	
13\h 14\m 08\fss645 &  36\degs 15\amin 43\farcs77 &   1.69 &   2.08 \\	
13\h 14\m 09\fss508 &  36\degs 28\amin 54\farcs98 &   0.85 &   1.35 \\	
13\h 14\m 17\fss867 &  36\degs 49\amin 14\farcs62 &  ^{4}15.91 &  16.20 \\	
13\h 14\m 29\fss894 &  36\degs 42\amin 55\farcs64 &   3.55 &   4.20 \\	
13\h 14\m 30\fss051 &  36\degs 23\amin 29\farcs51 &   0.79 &   1.15 \\	
13\h 14\m 39\fss929 &  36\degs 35\amin 04\farcs09 &   0.69 &   1.10 \\	
13\h 14\m 42\fss432 &  36\degs 39\amin 24\farcs84 &   0.93 &   0.74 \\	
13\h 14\m 43\fss392 &  36\degs 32\amin 18\farcs47 &   2.50 &   2.65 \\	
13\h 14\m 44\fss308 &  36\degs 39\amin 02\farcs59 & ^{1}226.11 & 228.37 \\	
13\h 15\m 25\fss329 &  36\degs 47\amin 13\farcs79 &   1.29 &   1.90 \\	
13\h 15\m 28\fss865 &  36\degs 50\amin 08\farcs62 &   0.95 &   1.80 \\	
\hline \noalign{\smallskip}
 \end{array}
$$
\begin{list}{}{}
\item[] {$^\dagger$ Background sources of the 1.4~GHz NGC~5033 field,
from VLA observations obtained on 7 August 1993 with the VLA-C. 
$S_{1.4}^P$ and $S_{1.4}^I$ 
are the peak and integrated flux densities
  of each component, respectively. We used the {\sc AIPS} task SAD
to search for point-like sources above a flux density level cutoff
of five times the off-source (1\,rms=93\muJybeam).
$^{1}$\citet{colla73}; 
$^{2}$\citet{mcmahon02}; 
$^{3}$\citet{ho01};
$^{4}$\citet{hales88}; 
}
\end{list}
\end{table}

\subsection{Background sources}

We obtained the 1.4~GHz VLA-C image in Fig.~1 (panel a) by using
wide-field techniques, to accurately take into account the
contribution to the total radio emission from confusing sources within
an area of $50\amin \times 50\amin$ around the nucleus of NGC~5033.
We used the {\sc AIPS} task SAD to search for point-like sources above
a flux density level cutoff of five times the off-source
(1\,rms=93\muJybeam).  This resulted in the detection of a total of 49
background sources (i.e., outside the lowest drawn contour in panel
(a) of Fig.~\ref{fig,vla}), detected above five times the off-source
rms (1 rms = 93\muJybeam).  In Table~\ref{tab,back}, we list the
coordinates, peak, and integrated flux densities for each source
component detected at 1.4~GHz. A number of sources exist in the
literature (see caption of Table~\ref{tab,back}), but many of them are
detected for the first time.  The most striking finding among these
background sources is B2~1312+36 \citep{colla73}, which showed a peak
flux density of 226.1\mJybeam on 7 August 1993, and of 4.6\mJybeam at
4.9~GHz on 8 April 2003. The enormous difference in flux density
between the two frequencies would imply an ultra steep spectrum for
B2~1312+36.  However, the non-simultaneity of these observations do
not allow for a precise estimate of $\alpha$. Fortunately, we were
able to make use of snapshot 1.4~GHz VLA observations taken also on 8
April 2003, to precisely determine $\alpha$.  The 1.4~GHz peak flux
density of B2~1312+36 was of 190.3\mJybeam, indicating
$\alpha_{1.4}^{4.9} = -3.1\pm 0.1$ ($S_\nu \propto \nu^\alpha$), and
thus confirming that B2~1312+36 is an ultra steep spectrum (USS)
source. Such steep spectrum is found either in pulsars, or in very
variables sources (e.g.  \citealt{debreuck00}, \citealt{izvekova81}).
A FIRST image of B2~1312+36, taken on 17 July 1994, shows a 1.4 GHz
peak flux density of 438.6~mJy/b, which confirms this source displays
strong flux density variations, and hence is very unlikely to be a
pulsar.  The source has no optical counterpart at the limit of the
POSS~I plates ($m_R\approx$20.0) and SDSS plates ($m_v\approx$24.0),
neither an infra-red counterpart at the limit of the 2MASS plates
(J$\approx$15.8; H$\approx$15.1; K$\approx$14.3).

\section{Discussion}
\label{sec,discussion}

\subsection{The radio spectrum and magnetic field of NGC~5033}
\label{sec,light-curve}

We used the VLA archival data obtained in 1999 under project AU0079 to
derive the spectral index of the core-jet structure that is seen in
panels (d) and (e) of Figure~\ref{fig,vla}.  Since the angular
resolution of the 4.9 GHz VLA observations in A-configuration is
essentially the same achieved by the 1.5~Ghz VLA observations in
B-configuration, we used the peaks of brightness at each frequency (as
given by the AIPS task IMSTAT) to estimate the spectral index of the
compact, nuclear source.  We obtained $\alpha = 0.44 \pm 0.04$, which
is in agreement with the spectral index published by \citet{ho01} and
confirms the AGN-like nature of the radio emission from the nuclear
region of this Sy~1.5 galaxy.  
Since the angular resolution of the 4.9~GHz VLA-D is very close to
that of the 1.4~GHz VLA-C, we can estimate the spectral index of the
kiloparsec scale disk of NGC~5033. Using the peaks of brightness at
those frequencies, we obtain $\alpha = -0.93\pm0.02 (S_\nu \propto
\nu^\alpha$, indicating an steep synchrotron spectral index of the
radio emission within the inner few kiloparsecs of NGC~5033.
\citealt{lisenfeld00} have shown that galaxies having steep spectral
radio indices ($\alpha \lsim -0.9$) with increasing distance from the
disk tend to have significant inverse Compton and synchrotron energy
losses of cosmic ray electrons, indicating that their escape rate is
low (e.g., M~82~\citep{seaquist91}; NGC~253~\citep{carilli92};
NGC~2146~\citep{lisenfeld96}; NGC~4666~\citep{sukumar88}).  While the
spectral index we find for the kiloparsec region of NGC~5033 hints to
the fact that energy losses are important, we cannot exclude that the
observed spectrum is due to a rather normal, relativistic electron
energy distribution with $p\approx2.8$ ($N(E) \propto E^{-p}$). 

The 1.4~GHz monochromatic luminosity of the core of NGC~5033, as
obtained with the VLA in A-configuration, is equal to of $L_{\rm 1.5}
= (1.2\pm0.2)\EE{27}\,D^2_{13}$\ergshz, and the integrated isotropic
radio luminosity between 300~MHz and 30~GHz is, for $\alpha=-0.44$, of
$L_{\rm core} = (1.5\pm0.2)\EE{37}\,D^2_{13}$\ergs.  From our VLA
observations in D-configuration, we obtain a 1.4~GHz monochromatic
luminosity for the nuclear regions of the galaxy NGC~5033 of $L_{\rm
  1.4} = (3.0\pm0.3)\EE{28}\,D^2_{13}$\ergshz.  If the spectral index
of $\alpha=-0.9$ extends at least from 300~MHz to 30~GHz, then the
integrated isotropic radio luminosity for the nuclear regions of the
galaxy is $L_{\rm R} = (2.3\pm0.2)\EE{38}\,D^2_{13}$\ergs.  We can then
estimate the average magnetic field in the radio emitting regions
NGC~5033.  Following \citet{pachol70}, we have

\begin{equation}
B_{\rm min} = (4.5\,c_{12}/\phi)^{2/7}\,(1 + \psi)^{2/7}\,  R^{-6/7}\, L_{\rm R}^{2/7}
\label{eq,bmin}
\end{equation}

\noindent
where $L_{\rm R}$ is the radio luminosity of the source, in \ergs;
$R$ is the characteristic linear size of the source, in cm;
$c_{12}$ is a slowly-varying function of the spectral index
\citep{pachol70}; $\phi$ is the filling factor of fields
and particles, i.e.  the ratio of the volume filled with relativistic
particles and magnetic fields to the total volume occupied by the
source; and $\psi$ is the ratio of the heavy particle energy to the
electron energy.  Since the value of $\phi$ is highly uncertain, we
will set it to $\phi$=0.5 for the sake of simplicity.  The value of
$\psi$ depends on the mechanism that generates the relativistic
electrons, and can range from 1 to 2000. We will adopt here a value of
$\psi = 100$, which seems appropriate for galaxies \citep{moffet73}. 
As the characteristic size of the source, $R$
we took an angular region of 30\asec$\times$30\asec centered in the nucleus
of NGC~5033, which approximately corresponds to the inner region of the 
galaxy disk. The corresponding approximate flux density values at each frequency
are of 58, 21, and 6~mJy at 1.4, 4.9, and 8.4~GHz.
With the above values, Eq.~\ref{eq,bmin} results in
minimum magnetic field values of 23, 22, and 18
$\mu{\rm G}$ at 1.4, 4.9, and 8.4~GHz, respectively. Therefore, 
if the situation is close to equipartition, the average magnetic field 
in the inner region of the disk is of about 20
$\mu{\rm G}$.

\subsection{Radio/FIR correlation and the star-formation rate of NGC~5033}
\label{sec,radio-fir}

The existing linear correlation between the total radio continuum
emission and the far-IR luminosity ($L_{\rm FIR}$) is well known in
``normal'' galaxies \citep{condon92}, and is
generally interpreted as being due to the presence of massive stars.  These
massive stars provide relativistic particles via supernova explosions
and heat the interstellar dust, which then re-radiates the energy at
FIR wavelengths (\citealt{helou85};
\citealt{condon92}).  The ratio of infrared to radio luminosity is
usually expressed through the $q$-parameter \citep{helou85}, 
which is defined as 
$q={\rm log}~ [(FIR/3.75 \times 10^{12})/S_{1.4~{\rm GHz}}]$,
where $S_{\rm 1.4 GHz}$ is the observed 1.4 GHz flux density in units
of W m$^{-2}$ Hz$^{-1}$,  and $FIR$ is given by
$
FIR=1.26 \times 10^{-14}(2.58S_{60~{\mu{\rm m}}}+S_{100~{\mu{\rm m}}})
$
where $S_{60\mu m}$ and $S_{100\mu m}$ are
IRAS 60 $\mu$m and 100 $\mu$m band flux densities, in Jy (Helou et
al.~1988).  

Therefore, $q$ is a measure of the logarithmic FIR/radio flux-density
ratio, and is an indicator of the relative importance of an AGN, or
starburst in a galaxy.  Most galaxies in the IRAS Bright Galaxy Sample
have $q \approx 2.34$, though some galaxies have smaller $q$ values
due to additional contributions from compact radio cores and radio
jets/lobes \citep{sanders96}.
The bulk of the far infra-red emission comes from a region of angular
size of radius 30\asec ($\approx 1.9$~kpc), as measured from an
archival ISO image of NGC~5033.  Hence, to compute $q$, we estimated
the 1.4~GHz radio emission from a circumnuclear region of the same
angular size, using the AIPS task JMFIT. The resulting value is of
about 58~mJy, which translates into an isotropic luminosity of $L_{\rm
  1.4} = 1.2\EE{21}$\whz.  The IR flux densities at 60 and
100~$\mu{\rm m}$ are $S_{60~{\mu{\rm m}}}=16.20$~Jy and
$S_{100~{\mu{\rm m}}}=50.23$~Jy, respectively \citep{sanders03}.  The
resulting q-parameter -after subtraction of the radio emission
contributed by the nucleus- is $q$=2.83.  \citet{yun01}
pointed out that values of $q$ less than 1.64 would indicate the radio
dominance of an AGN.  Thus, it appears that the radio nucleus of the Sy~1.5
galaxy NGC~5033 is of very low-luminosity, and the radio emission from its
circumnuclear region is dominated by starburst activity.  
We can characterize such a starburst by, e.g., following the
prescriptions by \citet{scoville97}, which characterize a starburst by
its total luminosity, $L$, Ly continuum production, $Q$, and
accumulated stellar mass, $M_\star$.  Those values are obtained as
power-law approximations of the lower and upper mass cutoffs for
stars, $m_l$ and $m_u$, the constant rate of star formation, $\mdot$,
and the burst timescale, $t_B$.  In particular, a starburst lasting
10~Myr and producing stars at a rate of 2.8\msunyr\ in the range $m_l
= 1$\,\msun\ and $m_u = 40$\,\msun, yields essentially the same
observed far-infrared luminosity of NGC~5033 ($L_{\rm FIR} =
1.2\EE{10}$\,\lsun), and results into a Ly continuum production of
$Q\approx3\EE{53} {\rm s^{-1}}$, and $M_\star \approx 9\EE{7} \msun$.
To estimate the supernova rate, we used a minimum mass for yielding
Type II supernovae of 8$\msun$.  The corresponding supernova rate was
of 0.045\,yr$^{-1}$, which corresponds to about 1 SN event every 22
yr.  Therefore, we find that a rather modest (both in time and
intensity) starburst scenario is able to satisfactorily account for
both the observed far infra-red and radio luminosity of NGC~5033.
(Interestingly, NGC~5033 has been host galaxy to at least three
supernovae in the last 60 yr: SN~1950C, SN~1985L, and SN~2001gd,
although the supernova rate we have inferred applies for the nuclear
and circumnuclear regions, where the main bulk of radio and FIR
emission come.)

\subsection{The nature of the non-nuclear radio sources in NGC~5033}
\label{sec,radio-spurs}

We now discuss the nature of the four sources (N1, SW1, SW2, and E1;
see Table~\ref{tab,vla}) detected in the VLA images well within the
radio contours of the 1.4~GHz VLA-C radio emission (panels (a) and (c) in 
Fig.~\ref{fig,vla}.

\subsubsection{Source N1}

N1 is catalogued in \citet{ho01} as a background source. N1 is also
detected in our 4.9 and 8.4 GHz images, taken 10 years later than the
1.4 GHz in Fig. 1. (see also Table~\ref{tab,vla}).  Its spectral
index between 4.9 and 8.4 GHz is $\alpha \approx-1.1$, and if we
assume no variability between 1993 and 2003, the integrated spectral
index from 1.4 to 8.4 GHz is $\alpha = -0.8\pm0.1$.  From our 8.4 GHz
VLA-D image, which yields the highest resolution, the coordinates of
N1 are $\alpha$=13\h13\m33\fss44$\pm$0\fss02,
$\delta$=36\degs37\amin28\farcs7$\pm$0\farcs2 (J2000.0), These
coordinates are coincident with those reported for FIRST
J131333.4+363728 \citep{becker95}, which suggests N1 being a
background source. Moreover, N1 is about 6\asec\ away from the closest
H~II regions, so it is likely to be unrelated to them. We therefore
conclude that N1 is a background source.  

\subsubsection{Source SW1}

SW1 had a 1.4~GHz peak flux density of 4.5~mJy on April 1993, and was
also detected at 4.9~GHz and 8.4~GHz (peaks of 0.66~mJy and 0.36~mJy,
respectively). The source was tentatively identified as a background
source by \citet{ho01}, at a flux density level of 0.3~mJy at 4.9~GHz.
The source therefore shows some variability, but the up-and-down
behaviour at 4.9~GHz is at odds with it being a radio supernova.  The
closest H~II region is about 5\asec away from the radio position
\citep{evans96}, which suggests it is unrelated to it.  Finally, the
source was detected by ROSAT, but there is no detection in the
optical.  We therefore suggest, as previous authors, that SW1 is also
a background source.

\subsubsection{Source E1}

E1, detected in our 8.4~GHz VLA-D observations (panel (c) in
Figure~\ref{fig,vla}), also coincides with the source J131330.1+363537
detected by \citet{ho01} with a flux density of $S_{4.9} \approx
0.20\pm0.06$~mJy, and which these authors tentatively identified as a
background source.  Our continuum 8.4~GHz radio observations indicate
an angular size for E1 of $\simeq\,$7.4\asec (corresponding to a linear
size of $\simeq\,470$~pc). The 8.4 GHz flux density of above
2\,$\sigma$ is $S_{8.4} = (123\pm 19)\mu$Jy.  We also used {\em
  XMM-Newton} archival data of NGC~5033 taken on 18 December 2002 as
part of an observing program to observe SN~2001gd, and refer the
reader to \citet{mapt05} for further technical details.  We analyzed
the X-ray data in a region ofz 9\arcsec in radius (approximately the
same size as the radio emitting region), and centered at the peak of
8.4~GHz radio emission.  The X-ray spectrum has 200$\pm$20 counts, and
is best fit by an optically thin thermal plasma with temperature $k\,T
= 2.3^{+1.6}_{-0.6}$~keV and column density $N_{\rm H} =
(1.8^{+1.3}_{-0.8}) \EE{21}$~cm$^{-2}$, for a galactic column density
of $N_{\rm H}^{\rm gal} = 1.1\EE{20}$~cm$^{-2}$.

For the reasons outlined in the previous paragraph, we suggest that E1
is a radio ''spur'' which emanates from the disk of NGC~5033.  Such
radio spurs can be observed as a consequence of the formation of
superbubbles of size $\sim300-1000$~pc, if sequential Type II
supernovae explosions occur every $\sim 10^5$~yr during the lifetime
of a giant molecular cloud, canonically taken to be $\sim 1-2
\EE{7}$~yr (\citealt{norman89} and references therein) and have been
previously detected, e.g., in NGC~253~\citep{heesen04}.  The energy
input by stellar winds and multiple supernovae resulting from OB
associations results in an overpressurized region (the bubble), that
expands and pops out of the disk. The supernovae explosions would then
accelerate the relativistic electrons that, immersed in a magnetic
field, would be responsible for the observed synchrotron radio
emission, which is observed as a radio spur.

The non-thermal spectral index for the spur, $\alpha_{4.9}^{8.4}
\simeq$-0.9, is similar to the non-thermal spectra observed for radio
supernovae.  If the synchrotron spectrum holds at least between
300~MHz and 30~GHz, the total radio luminosity of the spur would be
1.1$\EE{36}\,D^2_{13}$\ergs, a typical value for relatively old radio
supernovae, or young supernova remnants.  We can then estimate the
minimum total energy in relativistic particles and fields, and the
approximate equipartition magnetic field (Eq.~\ref{eq,bmin}) in the
radio spur from minimum energy arguments \citep{pachol70}:

\begin{equation}
E_{\rm min} = c_{13}\, (1 + \psi)^{4/7}\, \phi^{3/7}\, R^{9/7}\, L_{\rm R}^{4/7}
\label{eq,emin}
\end{equation}

\noindent
where $c_{13}$ is a slowly-varying function of the spectral index, 
and the rest of the parameters have the same meaning as in 
Eq.~\ref{eq,bmin}, and we will also use $\phi=0.5$ and $\psi=100$
for the sake of simplicity.

We then obtain
$ E_{\rm min}\,\simeq\,4\EE{52}\, {\rm ergs} $
and 
$ B_{\rm min}\,\simeq\, 13\, \mu{\rm G}$.
The lifetime of the radio spur can be obtained by assuming
that its observed radio luminosity has been constant. In that case,
$\tau_{\rm spur} = E_{\rm rel}/L_{\rm R}$. 
Since the relativistic particle energy is 
$E_{\rm rel} = 4/7 (1+\psi) E_{\rm min} \simeq 4.3 \EE{50}$~ergs, 
we get
$\tau_{\rm spur} \simeq$~16.1~Myr. 
This value agrees well with the expected lifetime for giant molecular
clouds, estimated to be $\sim10-20$~Myr.

The characteristic lifetime of electrons undergoing 
radiative synchrotron losses is (see, e.g., \citealt{pachol70})

\begin{equation}
\tau_{\rm syn} = 
\frac{E}{-(dE/dt)_{\rm syn}} \simeq
33.4\,B_{10}^{-3/2}\,\nu_1^{-1/2}\, {\rm Myr}
\label{eq,lifetime}
\end{equation}

\noindent
where $B_{10}$ is the magnetic field in units of 10$\mu$G, and
$\nu_{1}$ is the critical frequency, in GHz.  Hence, the non-thermal
electrons emitted in the radio spur have lifetimes of about 19.0,
8.6, and 6.6~Myr at 1.4, 4.9, and 8.4~GHz, respectively.  It then
follows that while the low-frequency radio emission (1.4~GHz) of the
radio spur may be due to synchrotron losses of the initially injected
electron population, the higher frequency radio emission requires
either subsequent injection of fresh electrons, or a
reacceleration of the old ones, or both. These facts are all in
agreement with the cosmic ray electrons in the radio spur being
injected by young supernovae and supernovae remnants occurring during
the lifetime of the host giant molecular cloud.

In summary, while we cannot completely exclude that N1 is a background radio
source, there is huge evidence for it being a galactic source, namely
a radio spur.  Indeed, the size ($\simeq 470$~pc), radio spectrum
($\alpha \simeq -0.9$), high temperature of the ionized plasma
($T_{\rm e} \simeq 2.5\EE{7}$~K), and radiative lifetime of the
putative spur ($\simeq 16$~Myr) suggest that it could be a hot
superbubble formed as a consequence of sequential supernova explosions
occurring during the lifetime of a giant molecular cloud.  

\subsubsection{Source SW2}

SW2 was detected at 1.4~GHz in August 1993 ($S_{1.4}^P =
1.85$\mJybeam), and further identified at 4.9~GHz in April 2003
($S_{4.9}^P = 0.32$\mJybeam).  The radio location of SW2 is spatially
coincident, within less than 1.5\asec, with an H~II region of NGC~5033
\citep{evans96} and an optical counterpart from the NOMAD catalogue
\citep{zacharias04} (observed in 1963 with $B=19.64$ and
$R=19.97$). In turn, this H~II region is close to a region detected
with the WFPC2 camera onboard HST, which suggests their emission is
physically related.  Thus, SW2 would appear to be of galactic origin.
A possible scenario could be that proposed for E1 in the previous
paragraphs, i.e., that SW2 is a radio spur. The angular size of the
spur, about 8\asec in radius ($\approx$500~pc) is a typical value for
the putative superbubble.  An alternative (galactic) scenario could be
that of a radio supernova that exploded around 1993. The observed
1.4~GHz luminosity would then correspond to an event about three times
fainter than SN~2001gd at its peak.  However, both the radio spur and
the radio supernova scenarios are very difficult to reconcile with the
fact that we did not detect any 8.4~GHz emission on April 2003 above
57$\mu$Jy (3\,$\sigma$) around the position of SW2, which implies a
very steep spectrum $\alpha_{4.9}^{8.4} \approx -3.0$, and which
rather suggests an extragalactic origin for this source. In sum, while
it would seem that SW2 belongs to NGC~5033, we cannot rule out the
possibility that this is a background source.
  
\section{Summary}
\label{summary}

We have presented continuum VLA observations of the Sy~1.5 galaxy
NGC~5033, made at 4.9 and 8.4 GHz on 8 April 2003, and also archival
VLA data at 1.4 and 4.9 GHz, which probe the radio emission of this
galaxy from regions of less than a hundred parsecs up to more
than 35~kpc in size. We summarize our main results as follows:

\begin{itemize}

\item
The high-resolution VLA images show a core-jet structure for the
Sy~1.5 nucleus. The core has a moderately steep radio spectrum,
($S_{\nu} \propto \nu^\alpha; \alpha_{1.5}^{4.9} = -0.44 \pm 0.04$).
The 1.4~GHz radio emission, as traced by the VLA in C configuration,
correlates very well with the optical emission of the whole galaxy,
and delineates exquisitely the spiral arms. Our 4.9 and 8.4~GHz
simultaneous observations with the VLA-D show only significant radio
emission from the inner disk of the galaxy. Combining the data at 1.4,
4.9, and 8.4 GHz, we find that the radio emission of the inner disk of
the galaxy shows a steep spectrum ($\alpha_{1.4}^{8.4} \approx -0.9$),
which is usually found in galaxies where radiative losses are high in the disk
of the galaxy, so that the escape rate of their cosmic ray electrons
is low \citep{lisenfeld00}. 

\item
If the synchrotron spectrum extends at least from 300~MHz to 30~GHz,
the isotropic radio luminosity of the inner disk of NGC~5033 is
$L_{\rm R}(2.9\pm0.3)\EE{21}\,D^2_{13}$~W\,m$^{-2}$\,Hz$^{-1}$.
Combining this value with the far-infrared luminosity of the galaxy
($L_{\rm FIR} = 1.20\EE{10}$\lsun), we obtained a value for the $q$
parameter of 2.83, which indicates that the radio emission from
the inner regions of NGC~5033 is mainly powered by a recent
starburst. In fact, the compact core-jet structure at the center of
the galaxy, contributes only about 7\% of the total 1.4~GHz radio
luminosity.  We find that a relatively short starburst, lasting 10~Myr
and and producing stars at a rate of 2.8\msunyr\ in the range $m_l =
1$\,\msun\ and $m_u = 40$\,\msun, reproduces very well the observed
radio and far-infrared luminosities, and results into a supernova rate
of 0.045\,yr$^{-1}$.  This supernova rate corresponds to about 1 SN
event every 22 yr. 

\item 
We find evidence for the existence of a radio ''spur'' (E1) of
radius $r \simeq 470$~pc, located at a distance of about 2.1~kpc above
the disk plane.  Its radio spectrum ($\alpha \simeq -1.1$), high
temperature of the ionized plasma ($T_{\rm e} \simeq 2.5\EE{7}$~K),
and lifetime ($\tau \simeq 16$~Myr) strongly suggest that it is a hot
superbubble formed as a consequence of sequential supernova explosions
occurring during the lifetime of a giant molecular cloud.

\end{itemize}

\section*{Acknowledgments}
We thank the referee, Marek Kukula, for his comments and careful
reading of the manuscript. We also thank Mart\'{\i}n Guerrero for
helping us with the X-ray calculations, and to Enrique P\'erez for
helpful discussions.  This research was partially funded by grants
AYA2005-08561-C03-02 and AYA2002-00897 of the Spanish Ministerio de
Ciencia y Tecnolog\'{\i}a.  MAPT is supported by the Spanish National
programme Ram\'on y Cajal.  NRAO is a facility of the USA National
Science Foundation operated under cooperative agreement by Associated
Universities, Inc. We made extensive use of the NASA Astrophysics Data
System Abstract Service and of the Aladin (v4.006) software, developed
and maintained by the Centre de Donn\'ees Astronomiques de Strasbourg
(CDS).

\bsp

\label{lastpage}

\end{document}